\documentclass[aps,prl,twocolumn,footinbib,superscriptaddress]{revtex4-1}
\usepackage[T2A]{fontenc}
\usepackage[utf8]{inputenc}
\usepackage{amsmath}
\usepackage{amssymb}
\usepackage{color,xcolor}
\usepackage{natbib}
\usepackage[colorlinks=true,bookmarks=false,citecolor=blue,urlcolor=blue]{hyperref} 
\usepackage{bm}
\usepackage{epsfig}
\usepackage{verbatim} 
\usepackage{morefloats}
\usepackage{graphicx}
\usepackage{bibentry}
\usepackage[mathscr]{euscript}
\usepackage{enumerate}
\usepackage{ bbold } 
\usepackage{ upgreek }
\usepackage{subfigure}
\usepackage{amsmath}
\usepackage{amssymb}

\usepackage{bm}

\renewcommand{\Re}{\mathrm{Re}}
\renewcommand{\Im}{\mathrm{Im}}

\def\dfrac{\displaystyle\frac}  

 \usepackage{lipsum}
 
\renewcommand{\phi}{\varphi}

\begin{document}

\title{Self-steepening-induced stabilization of nonlinear edge waves\\ at photonic valley-Hall interfaces}

\author{\firstname{Ekaterina O.} \surname{Smolina}}
\affiliation{Department of Control Theory, Nizhny Novgorod State University, Gagarin Av. 23, Nizhny Novgorod, 603950 Russia}

\author{\firstname{Lev A.} \surname{Smirnov}}
\affiliation{Department of Control Theory, Nizhny Novgorod State University, Gagarin Av. 23, Nizhny Novgorod, 603950 Russia}

\author{\firstname{Daniel} \surname{Leykam}}
\affiliation{Centre for Quantum Technologies, National University of Singapore, 3 Science Drive 2, Singapore 117543}

\author{\firstname{Daria A.} \surname{Smirnova}}
\affiliation{ARC Centre of Excellence for Transformative Meta-Optical Systems (TMOS), Research School of Physics,
Australian National University, Canberra, ACT 2601, Australia} 
\affiliation{Theoretical Quantum Physics Laboratory, Cluster for Pioneering Research, RIKEN, Wakoshi, Saitama 351-0198, Japan}

\begin{abstract}
Localized nonlinear modes at valley-Hall interfaces in staggered photonic graphene can be described in the long-wavelength limit 
by a nonlinear Dirac-like model including spatial dispersion terms. 
It leads to a modified nonlinear Schr\"odinger equation for the wave field amplitude 
that remarkably incorporates 
a nonlinear velocity term.
We show that this nonlinear velocity correction results in a counter-intuitive stabilization effect for relatively high-amplitude plane-wave-like edge states, which we 
confirm by calculation of complex-valued small-amplitude perturbation spectra and direct numerical simulation of 
propagation dynamics 
 in staggered 
 honeycomb waveguide lattices with on-site Kerr nonlinearity. Our findings are relevant to a variety of nonlinear photonic 
 systems described by Dirac-like Hamiltonians. 
\end{abstract}

\maketitle

\section{I. Introduction} 

Topological edge modes are the indicative hallmark of the topologically nontrivial systems that can be characterised by the quantised invariants of the bulk eigenspectrum. Driven by inspiration from the solid-state physics, they were 
observed in many engineered photonic platforms including waveguide arrays and photonic crystals~\cite{Ozawa2019}. Their dispersion can often be captured by effective Dirac-like models. 
Their   
transformations induced by nonlinear effects in optical systems constitute an important subject of research   
in pursuit of potential applications in high-speed photonic circuits and communication networks~\cite{Smirnova2020APR}. 

 \begin{figure}[b!]
	\centering
\includegraphics[width=8.1cm]{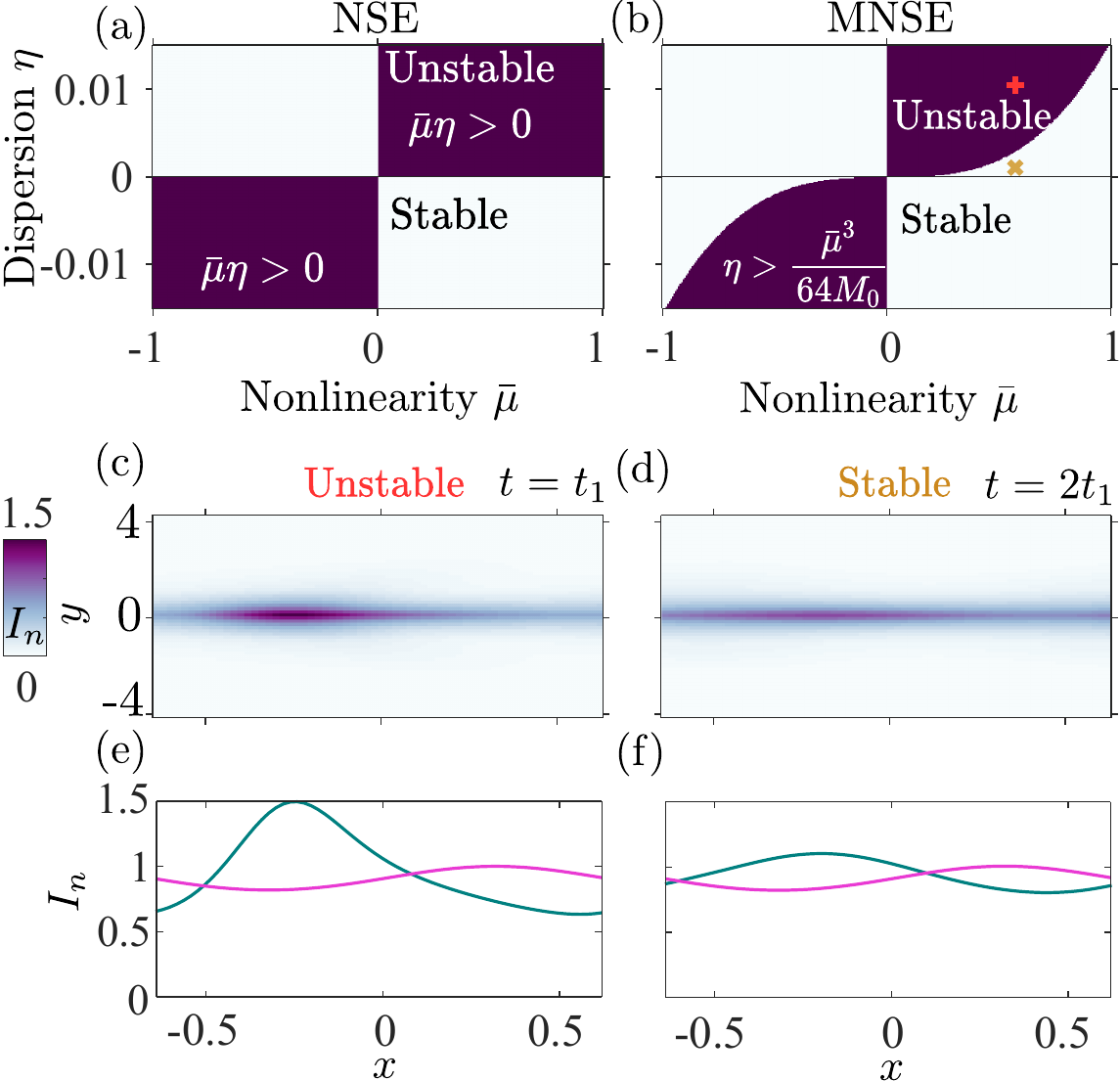}
\caption{
(a,b) Stability analysis based on the amplitude equation~\eqref{eq:solw}. Parameter plane of dispersion and nonlinearity illustrating 
stability (white) and instability (purple) areas for the propagating nonlinear edge modes, if the nonlinear velocity term is (a) omitted (conventional NSE), and (b) properly taken into account as written in MNSE Eq.~\eqref{eq:solw}. (c,d) The normalized intensity distribution denoted $I_n$ in the plane $(x,y)$ for (c, plus) unstable and (d, cross) stable edge waves numerically calculated in the framework of dynamic NDM~\eqref{eq:Dirac_main} at time moments $t_1=8$ and $2t_1$, respectively. Normalization is to the maximum value at the domain wall at the initial time. 
(e,f) Intensity profiles along the domain wall at the initial moment $t=0$ (pink lines) and after evolved (green lines) to the times corresponding to panels (c,d). Parameters are $M_0 = 1$, $g = 0.6$, $k=0$, the perturbation wave vector $\kappa=4.9$, $I_1 =1$, 
$\eta = ({\bar{\mu}^3}/{64 M_0})/3 $ (stable), 
$\eta = ({\bar{\mu}^3}/{64 M_0})\times 3 $ (unstable), where $ {\bar{\mu}} = (g I_1)/M_0$,  cf. Eq.~\eqref{eq:eta_unstable}.
}
\label{fig:mi1}
\end{figure}

Modulational instability is a phenomenon that appears 
in many nonlinear systems in nature as a result of the interplay between the nonlinearity and dispersion. In the course of this process development even minor disturbances to the stationary state in a nonlinear system experience exponential growth over time. 
For the boundary problem, 
it may in turn apply to the edge waves that propagate along the topological domain walls,; even when protected against backscattering these waves can become unstable under long-wavelength perturbations and break down into localized structures. Modulational instability can be used to probe bulk topological invariants~\cite{leykam2021probing,maluckov2022nonlinear,mancic2023band} and plays an important role in edge soliton formation~\cite{lumer2016instability,kartashov2016modulational,zhang2019interface,smirnova2021gradient}. 

Recent observations of optical solitons in Floquet topological lattices~\cite{Maczewsky2020,Mukherjee2020,Mukherjee2021} and other related phenomena, such as nonlinear Thouless pumping~\cite{Jrgensen2021,Jrgensen2023}, implemented in periodically modulated waveguide arrays reflect ongoing experimental interest in nonlinear effects in topological bands. 
It is argued that valley-Hall photonic lattices, being simple in design with no need for helical modulation, can be used to combine slow-light enhancement of nonlinear effects with topological protection against back reflection and disorder~\cite{Rechtsman2019,Sauer2020,Arregui2021}. Nevertheless, their performance versus conventional (non-topological) waveguides is still under debate and sensitive to the fabrication 
tolerance of the specific design implementation, as discussed in the recent experimental work Ref.~\cite{rosiek2023observation} demonstrating enhanced backscattering in valley-Hall 
photonic crystal slabs. 

Most previous studies~\cite{lumer2016instability,kartashov2016modulational,zhang2019interface} noted that the nonlinear counterparts of the topological edge modes in the optical systems with the self-focusing nonlinearity are modulationally unstable, and referred to the nonlinear Schr\"odinger equation (NSE) for the qualitative explanation~\cite{Ablowitz2013,Ablowitz2014,lumer2016instability,kartashov2016modulational,Ivanov2020,Ivanov2020b}. Here, we unravel the overlooked stabilization of the relatively high-amplitude nonlinear edge waves originating from the linear counterparts in the Dirac-like systems. It is rooted in the nonlinear velocity term correction to the NSE derived in Ref.~\cite{smirnova2021gradient}, which appears at interfaces between media with topological band gaps of finite width. The nonlinear velocity generally manifests itself in pulse self-steepening 
observable in experiments 
~\cite{anderson1983nonlinear,panoiu2009self,travers2011ultrafast,husko2015giant}. 
The instability inhibition at larger powers can loosely be interpreted as balanced compensation between slowly moving humps and faster moving drops. 

This paper begins by examining the linear stability of the nonlinear edge waves localized near domain walls
in the framework of the generic nonlinear Dirac equations, using purely analytical asymptotic analysis put forward in our earlier  works~\cite{smirnova2019topological,smirnova2021gradient}. We then proceed to the discrete lattice model based on the tight-binding description before finally presenting numerical modeling of a realistic optical implementation using optical waveguide arrays. These steps collectively  
constitute a comprehensive methodological set 
and self-consistently confirm the stabilization effect.  

\section{II. Continuum Nonlinear Dirac model} 
Our starting point is the nonlinear Dirac model (NDM) that describes that describes the spatiotemporal evolution of a two-component wavefunction ${\bf{\Psi}}=(\Psi_1,\Psi_2)^T$:
\begin{subequations} \label{eq:Dirac_main}
\begin{gather}
i\partial_t {\bf{\Psi}} \! = \hat{H} {\bf{\Psi}}, \\
\hat{H} = 
\left( \!{\begin{array}{*{22}{c}}
{M - g |\Psi_1|^2} &  {\hat{d}}\\
{\hat{d}^*}  & {- M - g |\Psi_2|^2}
\end{array}} \! \right),
\end{gather}
\end{subequations}
where the off-diagonal spatial derivative operator in the valley-Hall systems is defined as $\hat{d}={- i \partial_x -  \partial_y {- \eta \left( - i \partial_x +  \partial_y \right)^2}}$~\cite{smirnova2019topological,smirnova2021gradient}, and $t$ is the evolution coordinate, corresponding to propagation distance $z$ in case of waveguide arrays. 
Note, taking into account the second-order derivatives responsible for the spatial dispersion 
is significant 
for the correct description of the system behavior in the nonlinear regime, in particular, modulational instability, which is absent
in the NDM with $\eta=0$.

A topological domain wall is formally introduced by inverting the sign of the effective mass in two half-spaces, 
$M(y>0)=M_0,~M(y<0)=-M_0$. We take parameter $M_0 > 0$ without loss of generality. The work~\cite{smirnova2019topological} presents the analytical solution for the propagating nonlinear edge modes confined to the interface 
at $y=0$ and possessing the profiles $(\psi_1^0(y),\psi_2^0(y))^T e^{-i\omega_{\text{NL}}t + i k x}$ 
and nonlinear dispersion $\omega_{\text{NL}}(k,I_1) = - k - g I_1/2$. Here $I_1=|\psi_{1,2}(y=0)|^2$ is the intensity of this edge mode components at the interface. 
Although this formula for $\omega_{\text{NL}}(k,I_1)$ was derived at $\eta=0$, it is still applicable in the vicinity of $k=0$ for small $\eta$.

 In Ref.~\cite{smirnova2021gradient}, we investigated dynamics of edge wavepackets that vary slowly along the $x$ direction and derived the evolution equation for the slowly varying amplitude $a(t,\xi)$, where $\xi = x + t$ is a travelling coordinate, of edge pulses with accuracy order $\sim \mu^2$ (the small parameters are $g I_1/ 2 M_0 \sim \mu \ll 1,~\eta M_0 \sim \mu^2$): 
\begin{equation}
\label{eq:solw}
i\frac{\partial a}{\partial {t}} \approx - \dfrac{g}{4} |a|^2 a - i \dfrac{g^2}{32 M_0^2}  {|a|^2 \dfrac{\partial |a|^2} {\partial \xi} a} - \eta \dfrac{\partial^2 a} {\partial \xi ^2} + M_0^2 \eta a 
 \:.
\end{equation} 
It enters the asymptotic expression for the spinor components,  
\begin{equation} \label{eq:Psi12_series}
\Psi_{1,2} (x,y,t)  =  \pm   \dfrac{1}{\sqrt{2}} a (\xi; \left\{ \mu^n t \right\} )  e ^{-M_0 |y|} e^{ik\xi}+ 
\mathcal{O} (\mu) \:. 
\end{equation}
Equation~\eqref{eq:solw} differs from the usual  
nonlinear Schr\"odinger equation (NSE) due to the presence of a second higher-order nonlinear term (second term on the right hand of the equation), which accounts for phase modulation and self-steepening effects, and constitutes the nonlinear velocity; higher amplitude edge waves travel more slowly. As discussed in Ref.~\cite{smirnova2021gradient}, the nonlinear velocity term is a consequence of the asymmetric intensity-dependent localization of the edge states in the direction transverse to the interface.

 \begin{figure}[t!]
	\centering
\includegraphics[width=8.4cm]{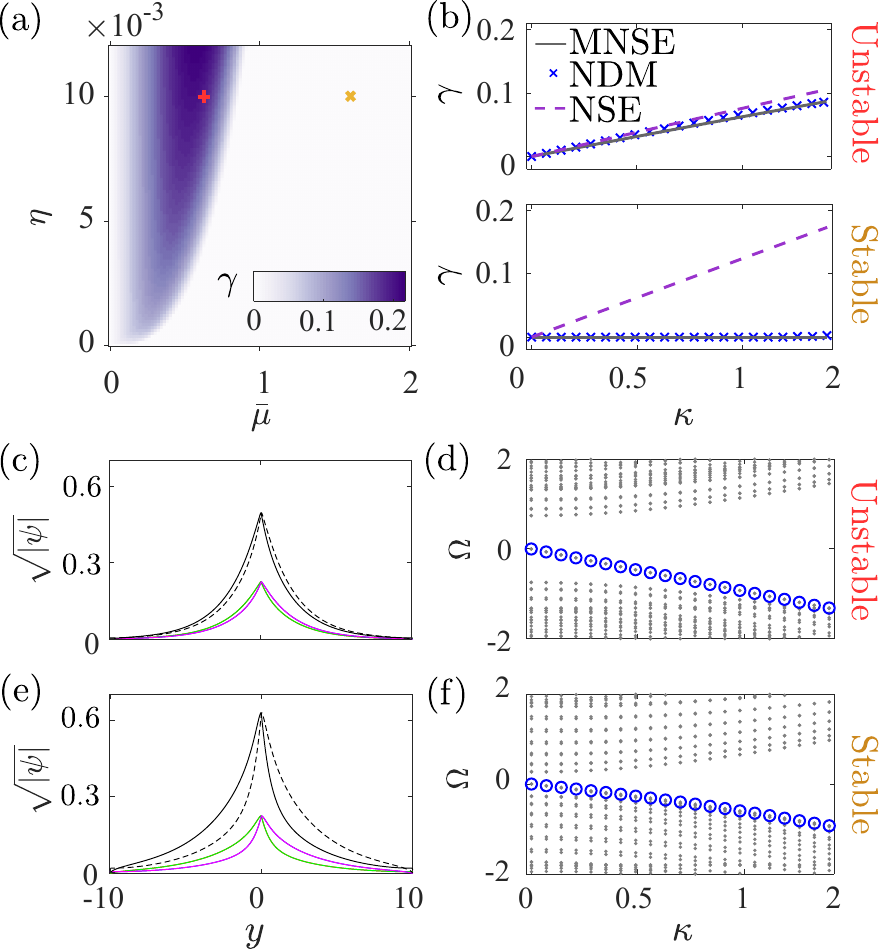}
\caption{
Linearised stability analysis based on Eqs.~\eqref{eq:Dirac_main},~\eqref{eq:solw}.
(a) Color map of the growth rate of instability in the parameter plane, calculated using Eq.~\eqref{eq:growth_rate}. 
(b) Comparison of the growth rate for transversely localized perturbations in stable and unstable cases obtained within NDM~\eqref{eq:Dirac_main} (blue crosses), MNSE~\eqref{eq:solw} (black solid line) and NSE (dashed purple line) disregarding the additional nonlinear velocity term in Eq.~\eqref{eq:solw}.
Field profiles (c,e) and (d,f) the real part of the perturbation eigenfrequency $\Omega \equiv \Re{\lambda }$ in (c,d) stable and (e,f) unstable cases, calculated using Eq.~\eqref{eq:Dirac_main}. Grey dots correspond to bulk perturbation bands. (c,e) Profiles of the spinor components in the nonlinear edge mode  ${\psi}^{\text{s}}_1(y)$ (black solid) and  ${\psi}^{\text{s}}_2(y)$ (grey dashed). Color lines: profiles of the perturbation eigenvector $\delta \psi_1$ (green) and $\delta \psi_2$ (pink).
Parameters are $M_0=1$, $g=10$, $\eta=0.01$, 
amplitudes $\sqrt{I_1}=0.25$ (unstable), $\sqrt{I_1}=0.4$ (stable). 
The magnitude square root in (c,e) is taken for better visualization of the different amplitude profiles within the same axis limits.}
\label{fig:mi2}
\end{figure}

Based on Eq.~\eqref{eq:solw}, the nonlinear edge wave's complex amplitude at the domain wall $y=0$ is given by 
\begin{equation} \label{eq:nonl_es_appr}
a = \mathcal{A}_0 e^{{i \left({g \mathcal{A}^2_0}/{4} - \eta   M_0^2\right)t}}\:,
\end{equation}
being exactly the steady state of this equation. 
In order to analyze stability of this state we apply a standard linear stability analysis by representing the perturbed solution in the form
\begin{equation} 
a\! = \! \left(  \mathcal{A}_0 \! + \! \delta  a_1 e^{-i\lambda t + i \kappa \xi} \! + \! \delta  a_2^* e^{i\lambda^* t - i \kappa \xi} \right) e^{{i \left({g \mathcal{A}^2_0}/{4} -  \eta   M_0^2\right)t}}\:. \label{eq:perturbed}
\end{equation}
Here $\delta a_{1,2}$ are small perturbations. 
The eigenfrequency $\lambda=\Omega+i\gamma$ is a complex number obtained by solving the linear eigenvalue problem for the  perturbations in the first order of accuracy upon substituting Eq.~\eqref{eq:perturbed} into Eq.~\eqref{eq:solw}. 
The resulting dependence of $\lambda$ on the 
modulation wavenumber $\kappa$ determines stability of the edge state, with $\gamma \equiv \Im(\lambda)$ being the growth rate. If $\gamma<0$, the nonlinear state is stable and only exhibits small amplitude oscillations in the presence of perturbations. However, if $\gamma>0$, the nonlinear state is unstable, resulting in significant profile variations during propagation. 

In this way, the growth rate is deduced to be 
\begin{equation} \label{eq:growth_rate}
\gamma(\kappa) = - i \dfrac{\bar{\mu}^2 \kappa} {8}  \pm \sqrt{\eta \kappa^2 \Bigl( \bar{\mu}  M_0  - \eta \kappa^2 \Bigr) - \dfrac{\bar{\mu}^4 \kappa^2}{64}} \:,
\end{equation}
where we denote the parameter of nonlinearity $\bar{\mu} = g I_1 / M_0 \equiv g \mathcal{A}_0^2 / 2 M_0$.  At $\eta=0$, i.e., in the absence of dispersion, the nonlinear edge wave is stable. The positive radicand $ \eta \kappa^2 \left( \bar{\mu} M_0  - \eta \kappa^2 \right) > {\bar{\mu}^4 \kappa^2}/{64}$ indicates instability. 
The instability condition can be formulated as follows,  
\begin{equation}\label{eq:eta_unstable}
 \eta >\dfrac{\bar{\mu}^3}{64 M_0} 
 = \dfrac{1}{M_0^4} \left( \dfrac{ g I_1}{4} \right)^3 .   
\end{equation}
This analysis notably reveals the counter-intuitive finding that large-amplitude waves can be stable for the parameters obeying~Eq.~\eqref{eq:eta_unstable}. Thus, once the nonlinear velocity is included into the modified nonlinear Schr\"odinger equation (MNSE), the instability area is reduced compared to the conventional NSE 
overlooking this contribution, as illustrated in Fig.~\ref{fig:mi1}(a,b). This means that the nonlinear velocity term has a stabilizing effect on the edge mode, making it less prone to decay. Examples of time evolution for modulationally unstable and stable edge waves modeled in the framework of NDM are shown in Figs.~\ref{fig:mi1}(c-f). 

To verify our analytical predictions, we further
calculate the perturbation spectra in Eq.~\eqref{eq:Dirac_main} directly.
To this end, we follow the procedure similar to the described above for Eq.~\eqref{eq:solw} and analyse  
small perturbations to the numerically found nonlinear stationary solution 
$({\psi}^{\text{s}}_1(y), {\psi}^{\text{s}}_2(y))^T e^{ikx-i\omega_{\text{s}} t}$ at $k=0$. 
The transverse profiles of the nonlinear mode components are visualized in Fig.~\ref{fig:mi2}(c,e) in black color. They exhibit noticeable asymmetry with the respect to the domain wall in the higher-intensity stable wave. We substitute the functions $\Psi_{1,2} = \bigl({\psi}^{\text{s}}_{1,2}(y)+\delta \psi_{1,2}(x,y,t) \bigr) e^{- i \omega_{\text{s}} t}$ into~Eq.\eqref{eq:Dirac_main} assuming the modulation of the form $\delta \psi_{1,2}(x,y,t)=\delta \phi_{1,2}(y) e^{i \kappa x - i \lambda t} + \delta \tilde{\phi}_{1,2}^*(y) e^{-i \kappa x + i \lambda^* t}$.
The obtained spectrum of localized near domain wall linear perturbations is depicted in Fig.~\ref{fig:mi2}. 
We then fix the dispersion parameter 
$\eta$ and consider two different nonlinearity strengths $\bar{\mu}$ corresponding to stable and unstable scenarios. 
As seen, the MNSE and Eq.~\eqref{eq:growth_rate} provides a more accurate approximation of the growth rate for unstable cases than NSE. Moreover, the stabilization effect is observable only in the framework of MNSE, while entirely absent in the conventional NSE. Note, however, we can correctly predict the growth rate $\gamma$ 
until the real part of perturbation's frequency, undergoing the nonlinearity-caused shift, crosses the bulk band. At that point, the approximate analytical approach breaks down, since the perturbations are no longer localized near the domain wall.

\section{III. Staggered graphene lattice model}

\begin{figure}[t!]
	\centering
\includegraphics[width=8.4cm]{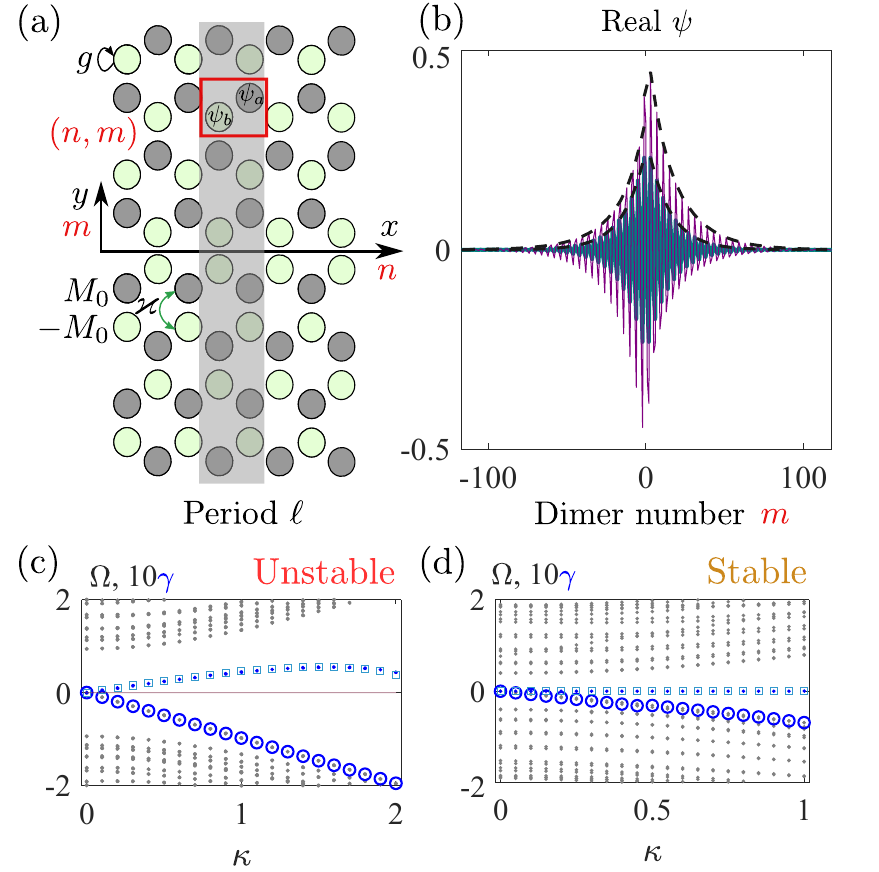}
\caption{Modulational instability in the tight-binding model.
(a) Schematic of a honeycomb lattice stripe with staggered sublattice potential $M(n,m)$ ($|M|=M_0$), which 
inverts the sign at a valley-Hall domain wall positioned at $y=0$. Here $\varkappa$ denotes the tunneling coefficient between elements within the tight-binding approximation, $\ell$ is the spatial period along horizontal axis, the nonlinear self-action effect is denoted by the circular arrow $g$.
(b) The real component of the field in the nonlinear edge state in the purely linear ($g=0$, green) and nonlinear (purple) regimes. Black dashed curves depict the field envelopes reconstructed in the continuum limit from the analytical model of Sec.~II.  
(c,d) The real $\Omega$ (gray encircled dots) and imaginary $\gamma$ (blue dots) parts of the eigenvalue for the localized near domain wall perturbations in the unstable (c) and stable (d) cases. Blue squares: the theoretical result Eq.~\eqref{eq:growth_rate}. Parameters are  $M_0=1$, 
$\varkappa = 7$, $g = 5$, amplitudes $\sqrt{I_1}=0.15$ (unstable), $\sqrt{I_1}=0.5$ (stable).
}
\label{fig:mi3}
\end{figure}

Given that staggered graphene~\cite{Ni2018,RingSoliton2018,smirnova2019topological,Smirnova2020LiSA,smirnova2021gradient} can be well-described by the NDM Eq.~\eqref{eq:Dirac_main} in the continuum limit, we will use 
a dimerized honeycomb lattice 
for further validation of our results with the example of 
a discrete 
two-dimensional system made of coupled sites.
We consider the ribbon geometry of the lattice, which is periodic along the horizontal ($x$) direction and has a finite size in the vertical ($y$) direction and utilize the tight-binding equations governing the propagation dynamics, which
assumes that each element is subject to linear interactions with the coupling coefficient $\varkappa$ with its' three nearest neighbours only: 
\begin{widetext} 
\begin{subequations} \label{eq:TBM} 
\begin{gather}
i \partial_{t} \psi_{a}(m, n)=M(m) \psi_{a}(m, n)-\varkappa(\psi_{b}(m, n)+\psi_{b}(m-1, n)+  \\ \nonumber
+0.5\left[\left(1+(-1)^{m}\right) \psi_{b}(m, n+1)+\left(1-(-1)^{m}\right) \psi_{b}(m, n-1)\right]) 
-g | \psi_{a}(m, n)|^2  \psi_{a}(m, n)\:,\\ 
i \partial_{t} \psi_{b}(m, n)=-M(m) \psi_{b}(m, n)-\varkappa (\psi_{a}(m, n)+\psi_{a}(m+1, n)+ \\ \nonumber
+0.5\left[\left(1+(-1)^{m}\right) \psi_{a}(m, n-1)
+0.5 \left(1-(-1)^{m}\right) \psi_{a}(m, n+1)\right]) 
- g | \psi_{b}(m, n)|^2  \psi_{b}(m, n),
\end{gather}
\end{subequations}
\end{widetext} 
where a pair of integers $(m, n)$ enumerates the dimer along $x$ (as $n$) and $y$ (as $m$) directions [see Fig.~\ref{fig:mi3}(a)], indices $a,b$ 
distinguish two different sublattices, and we introduced the local on-site nonlinerity of the strength $g$.
We consider the periodic stripe along $x$-direction, implying that the steady solution has the form $\psi_{a,b}(m, n)= \psi_{a,b}(m, K)  e^{iKn\ell-i\omega_{\text{s}} t}$. The period $\ell$ is chosen such that the Dirac velocity, being the coefficient in front of the first derivative, in the corresponding continuum NDM Eq.~\eqref{eq:Dirac_main}, is 
unity. In fact, NDM~\eqref{eq:Dirac_main} can readily be derived from the system~\eqref{eq:TBM} by expanding the Hamiltonian near the high-symmetry point $K=K_+={4\pi}/{3 \ell}$~\cite{Ni2018,Smirnova2020LiSA}, and, following this procedure, the dispersion coefficient is $\eta=1/6\varkappa$.

We search for the solution of system~\eqref{eq:TBM} in the form
\begin{equation}
\psi_{a,b}(m, n ) \!  = \!  \bigl( \psi_{a,b}(m, K) \! + \!  \phi_{a,b}(m, n) \bigr)  e^{iKn\ell-i\omega_{\text{s}} t}  ,
\end{equation}
where the wavefunction $\psi_{a,b}(m, K)$ represents the precise shape of the nonlinear Dirac edge mode [see Fig.~\ref{fig:mi3} (b)], which can be numerically obtained using Newton's method. On the other hand, $\phi_{a,b}(m, n)$ are small disturbances of the edge mode.
Similar to 
Sec.~II, we examine the eigenvalue spectra of the plane-wave-like perturbations expressed as $ \phi_{a,b}(m, n)= \delta \phi_{a,b} (m) e^{i \kappa n \ell - i \lambda t} + \delta \tilde{\varphi}_{a,b}^*(m) e^{-i \kappa n  \ell+ i \lambda^* t}$.
The results 
summarised in Fig.~\ref{fig:mi3} (c,d) are fully consistent with our findings in Sec.~II, 
signalling the presence of the nonlinear correction in MNSE.

\begin{figure*}[t!]
{\includegraphics[width=1\textwidth]{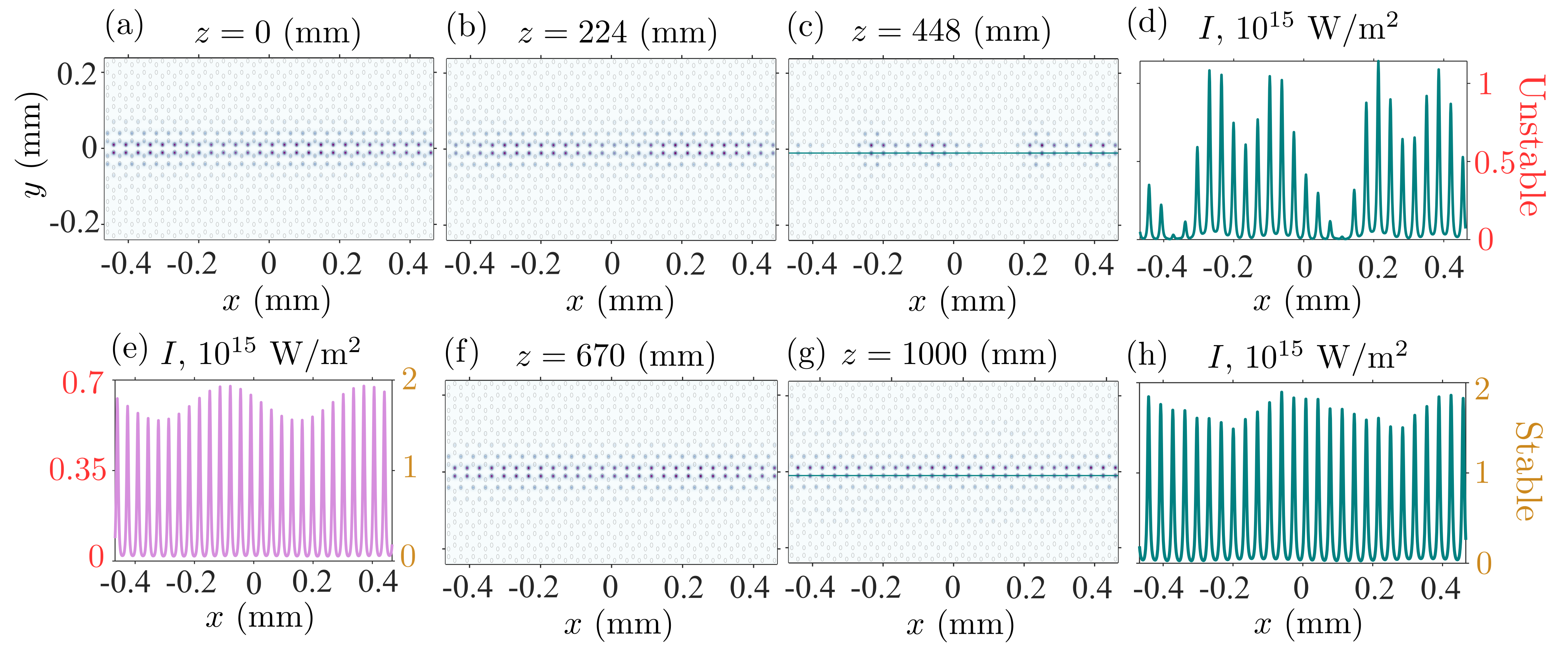}}
\caption{
Modulational instability of the nonlinear edge modes localized at the valley-Hall domain wall of the zigzag shape in an  optical honeycomb lattice of laser-written waveguides modelled in the paraxial approximation. The 
snaphshots of the intensity distributions  
(a,b,c) taken at different propagation distances indicate the instability development for the wave with smaller initial amplitude set in the input $z=0$, while the wave of the larger initial amplitude is stabilized (f,g). The cut along the domain wall (e) shows a profile of the initial excitation at $z=0$ and its transformation after the propagation at finite distance $z=448$~mm in (d) for the unstable edge wave, and at $z=1000$~mm in (h) for the stable edge wave.
}
\label{fig:mi4}
\end{figure*}

\section{IV. Optical implementation}

The discussed model can potentially be implemented in a range of settings, including optical lattices and metamaterials.
In this section, as a possible experimental 
platform, we examine valley-Hall waveguide arrays made of laser-written single-mode waveguides with parameters similar to those utilized in the experimental work Ref.~\cite{Noh2018}. The designed photonic lattice can be 
well described 
by the tight-binding model 
with effective parameters $\varkappa \approx 3 ~\mathrm{cm}^{-1}$ and $M_0 \approx 1~ \mathrm{cm}^{-1}$.
To study the 
evolution dynamics in the realistic array, we apply numerical techniques to solve 
Maxwell's wave equations in the paraxial approximation, namely, plane wave expansion to get the edge mode profile transverse to the interface and the beam propagation method to simulate propagation. 

As an initial condition, we set the plane-wave-like edge mode, whose amplitude is perturbed by 5\% large-scale noise [see Figs.~\ref{fig:mi4}(a,e)]. Then we directly model its dynamics up to large propagation distances along $z$ axis, 
being the evolution coordinate analogous to variable $t$ in Eq.~\eqref{eq:Dirac_main}, for the two different wave amplitudes falling into the unstable and stable regions in the parameter space for comparison. Representative snapshots of instability development are shown in Figs.~\ref{fig:mi4} (b,c,d).
The unstable edge state disintegrates 
into a series of soliton-like localized distributions.
On the contrary, in 
Figs.~\ref{fig:mi4}(f,g,h) we observe that the 
perturbed edge mode remains unchanged up to large distances, 
thereby indicating the stabilization effect.
Note, however, apart from the perturbations localized near the domain wall and well described by Eq.~\eqref{eq:solw}, the amplitude of the nonlinear wave in a realistic lattice can also undergo fluctuations caused by the bulk perturbations or coupling with other interfaces. This can shift a transition towards either the unstable or stable regime.

\section{Conclusion}

The performed study emphasises the importance of the nonlinear velocity term 
in the modified nonlinear Schr\"odinger equation for the adequate effective description of the nonlinear dynamics of edge waves 
supported by topological interfaces in the long-wavelength limit. 
As its' subtle consequence, the effect of the nonlinear edge mode stabilization at the valley-Hall interfaces was confirmed by the linearized 
stability analysis 
and direct dynamic modeling. 
Given the generality of the models and methods employed, our results establish the useful analytic concept of intuition for understanding dynamic effects in nonlinear topological photonic systems of various nature. 

\section{Acknowledgements}
E.S. and L.S. are supported in part by the MSHE under project No. 0729-2021-013. E.S. thanks the Foundation for the Advancement of Theoretical Physics and Mathematics "BASIS" (Grant No. 22-1-5-80-1). D.L. acknowledges support from the National Research Foundation, Singapore and A*STAR under its CQT Bridging Grant. D.S. acknowledges support from the Australian Research Council (DE190100430, CE200100010) and the Japan Society for the Promotion of Science under the Postdoctoral Fellowship Program for Foreign Researchers. 


\begin{thebibliography}{10}
\newcommand{\enquote}[1]{``#1''}

\bibitem{Ozawa2019}
T.~Ozawa, H.~M. Price, A.~Amo, N.~Goldman, M.~Hafezi, L.~Lu, M.~C. Rechtsman,
  D.~Schuster, J.~Simon, O.~Zilberberg, and I.~Carusotto, \enquote{Topological
  photonics,} Rev. Mod. Phys. \textbf{91}, 015006 (2019).

\bibitem{Smirnova2020APR}
D.~Smirnova, D.~Leykam, Y.~Chong, and Y.~Kivshar, \enquote{Nonlinear
  topological photonics,} Applied Physics Reviews \textbf{7}, 021306 (2020).

\bibitem{leykam2021probing}
D.~Leykam, E.~Smolina, A.~Maluckov, S.~Flach, and D.~A. Smirnova,
  \enquote{Probing band topology using modulational instability,} Physical
  Review Letters \textbf{126}, 073901 (2021).

\bibitem{maluckov2022nonlinear}
A.~Maluckov, E.~Smolina, D.~Leykam, S.~G{\"u}ndo{\u{g}}du, D.~G. Angelakis, and
  D.~A. Smirnova, \enquote{Nonlinear signatures of {F}loquet band topology,}
  Physical Review B \textbf{105}, 115133 (2022).

\bibitem{mancic2023band}
A.~Mancic, D.~Leykam, and A.~Maluckov, \enquote{Band relaxation triggered by
  modulational instability in topological photonic lattices,} Physica Scripta
  (2023).

\bibitem{lumer2016instability}
Y.~Lumer, M.~C. Rechtsman, Y.~Plotnik, and M.~Segev, \enquote{Instability of
  bosonic topological edge states in the presence of interactions,} Physical
  Review A \textbf{94}, 021801 (2016).

\bibitem{kartashov2016modulational}
Y.~V. Kartashov and D.~V. Skryabin, \enquote{Modulational instability and
  solitary waves in polariton topological insulators,} Optica \textbf{3},
  1228--1236 (2016).

\bibitem{zhang2019interface}
Y.~Zhang, Y.~V. Kartashov, and A.~Ferrando, \enquote{Interface states in
  polariton topological insulators,} Physical Review A \textbf{99}, 053836
  (2019).

\bibitem{smirnova2021gradient}
D.~A. Smirnova, L.~A. Smirnov, E.~O. Smolina, D.~G. Angelakis, and D.~Leykam,
  \enquote{Gradient catastrophe of nonlinear photonic valley-{H}all edge pulses,}
  Physical Review Research \textbf{3}, 043027 (2021).

\bibitem{Maczewsky2020}
L.~J. Maczewsky, M.~Heinrich, M.~Kremer, S.~K. Ivanov, M.~Ehrhardt,
  F.~Martinez, Y.~V. Kartashov, V.~V. Konotop, L.~Torner, D.~Bauer, and
  A.~Szameit, \enquote{Nonlinearity-induced photonic topological insulator,}
  Science \textbf{370}, 701--704 (2020).

\bibitem{Mukherjee2020}
S.~Mukherjee and M.~C. Rechtsman, \enquote{Observation of {F}loquet solitons in
  a topological bandgap,} Science \textbf{368}, 856--859 (2020).

\bibitem{Mukherjee2021}
S.~Mukherjee and M.~C. Rechtsman, \enquote{Observation of unidirectional
  solitonlike edge states in nonlinear floquet topological insulators,} Phys.
  Rev. X \textbf{11}, 041057 (2021).

\bibitem{Jrgensen2021}
M.~J\"{u}rgensen, S.~Mukherjee, and M.~C. Rechtsman, \enquote{Quantized
  nonlinear {T}houless pumping,} Nature \textbf{596}, 63--67 (2021).

\bibitem{Jrgensen2023}
M.~J\"{u}rgensen, S.~Mukherjee, C.~J\"{o}rg, and M.~C. Rechtsman,
  \enquote{Quantized fractional {T}houless pumping of solitons,} Nature Physics
  \textbf{19}, 420--426 (2023).

\bibitem{Rechtsman2019}
J.~Guglielmon and M.~C. Rechtsman, \enquote{Broadband topological slow light
  through higher momentum-space winding,} Phys. Rev. Lett. \textbf{122}, 153904
  (2019).

\bibitem{Sauer2020}
E.~Sauer, J.~P. Vasco, and S.~Hughes, \enquote{Theory of intrinsic propagation
  losses in topological edge states of planar photonic crystals,} Phys. Rev.
  Research \textbf{2}, 043109 (2020).

\bibitem{Arregui2021}
G.~Arregui, J.~Gomis-Bresco, C.~M. Sotomayor-Torres, and P.~D. Garcia,
  \enquote{Quantifying the robustness of topological slow light,} Phys. Rev.
  Lett. \textbf{126}, 027403 (2021).

\bibitem{rosiek2023observation}
C.~A. Rosiek, G.~Arregui, A.~Vladimirova, M.~Albrechtsen, B.~Vosoughi~Lahijani,
  R.~E. Christiansen, and S.~Stobbe, \enquote{Observation of strong
  backscattering in valley-hall photonic topological interface modes,} Nature
  Photonics p.~1 (2023).

\bibitem{Ablowitz2013}
M.~J. Ablowitz, C.~W. Curtis, and Y.~Zhu, \enquote{Localized nonlinear edge
  states in honeycomb lattices,} Phys. Rev. A \textbf{88}, 013850 (2013).

\bibitem{Ablowitz2014}
M.~J. Ablowitz, C.~W. Curtis, and Y.-P. Ma, \enquote{Linear and nonlinear
  traveling edge waves in optical honeycomb lattices,} Phys. Rev. A
  \textbf{90}, 023813 (2014).

\bibitem{Ivanov2020}
S.~K. Ivanov, Y.~V. Kartashov, A.~Szameit, L.~Torner, and V.~V. Konotop,
  \enquote{Vector topological edge solitons in {F}loquet insulators,} ACS
  Photonics \textbf{7}, 735--745 (2020).

\bibitem{Ivanov2020b}
S.~K. Ivanov, Y.~V. Kartashov, L.~J. Maczewsky, A.~Szameit, and V.~V. Konotop,
  \enquote{Bragg solitons in topological {F}loquet insulators,} Opt. Lett.
  \textbf{45}, 2271--2274 (2020).

\bibitem{anderson1983nonlinear}
D.~Anderson and M.~Lisak, \enquote{Nonlinear asymmetric self-phase modulation
  and self-steepening of pulses in long optical waveguides,} Physical Review A
  \textbf{27}, 1393 (1983).

\bibitem{panoiu2009self}
N.~C. Panoiu, X.~Liu, and R.~M. Osgood~Jr, \enquote{Self-steepening of
  ultrashort pulses in silicon photonic nanowires,} Optics letters \textbf{34},
  947--949 (2009).

\bibitem{travers2011ultrafast}
J.~C. Travers, W.~Chang, J.~Nold, N.~Y. Joly, and P.~S.~J. Russell,
  \enquote{Ultrafast nonlinear optics in gas-filled hollow-core photonic
  crystal fibers,} JOSA B \textbf{28}, A11--A26 (2011).

\bibitem{husko2015giant}
C.~Husko and P.~Colman, \enquote{Giant anomalous self-steepening in photonic
  crystal waveguides,} Physical Review A \textbf{92}, 013816 (2015).

\bibitem{smirnova2019topological}
D.~A. Smirnova, L.~A. Smirnov, D.~Leykam, and Y.~S. Kivshar,
  \enquote{Topological edge states and gap solitons in the nonlinear {D}irac
  model,} Laser \& Photonics Reviews \textbf{13}, 1900223 (2019).

\bibitem{Ni2018}
X.~Ni, D.~Smirnova, A.~Poddubny, D.~Leykam, Y.~Chong, and A.~B. Khanikaev,
  \enquote{$\mathcal{PT}$ phase transitions of edge states at $\mathcal{PT}$
  symmetric interfaces in non-{H}ermitian topological insulators,} Phys. Rev. B
  \textbf{98}, 165129 (2018).

\bibitem{RingSoliton2018}
A.~N. Poddubny and D.~A. Smirnova, \enquote{Ring {D}irac solitons in nonlinear
  topological systems,} Phys. Rev. A \textbf{98}, 013827 (2018).

\bibitem{Smirnova2020LiSA}
D.~Smirnova, A.~Tripathi, S.~Kruk, M.-S. Hwang, H.-R. Kim, H.-G. Park, and
  Y.~Kivshar, \enquote{Room-temperature lasing from nanophotonic topological
  cavities,} Light: Science {\&} Applications \textbf{9} (2020).

\bibitem{Noh2018}
J.~Noh, S.~Huang, K.~P. Chen, and M.~C. Rechtsman, \enquote{Observation of
  photonic topological valley {H}all edge states,} Phys. Rev. Lett.
  \textbf{120}, 063902 (2018).

\end{thebibliography}

\end{document}